\documentclass[twocolumn,showpacs,amsmath,amssymb,prl,aps,floats]{revtex4}
\usepackage{bm}% bold math
\usepackage{graphicx}
\usepackage{times}
\newcommand{\BoldVec}[1]{\mathchoice%
  {\mbox{\boldmath $\displaystyle     #1$}}%
  {\mbox{\boldmath $\textstyle        #1$}}%
  {\mbox{\boldmath $\scriptstyle      #1$}}%
  {\mbox{\boldmath $\scriptscriptstyle#1$}}%
}
\newcommand{\EQ}{\begin{equation}}
\newcommand{\EN}{\end{equation}}
\newcommand{\EQA}{\begin{eqnarray}}
\newcommand{\ENA}{\end{eqnarray}}

\newcommand{\Eq}[1]{Eq.~(\ref{#1})}

\newcommand{\bra}[1]{\langle #1\rangle}

\newcommand{\meanB}{\overline{B}}

\newcommand{\meanF}{\overline{\cal F}}

\newcommand{\meanBB}{\overline{\mbox{\boldmath $B$}}}
\newcommand{\meanJJ}{\overline{\mbox{\boldmath $J$}}}

\newcommand{\meanUU}{\overline{\mbox{\boldmath $U$}}}

\newcommand{\meanemfs}{\overline{\cal E} {}}
\newcommand{\meanemf}{\overline{\mbox{\boldmath ${\cal E}$}} {}}
\newcommand{\meanFF}{\overline{\mbox{\boldmath ${\cal F}$}} {}}
\newcommand{\calFF}{\overline{\mbox{\boldmath ${\cal F}$}} {}}
\newcommand{\kunit}{\hat{\mbox{$k$}} {}}

\newcommand{\rrr}{\BoldVec{r} {}}
\newcommand{\xx}{\BoldVec{x}{}}

\newcommand{\uu}{\BoldVec{u} {}}

\newcommand{\RR}{\BoldVec{R} {}}
\newcommand{\UU}{\BoldVec{U} {}}
\newcommand{\bb}{\BoldVec{b} {}}

\newcommand{\BB}{\BoldVec{B} {}}

\newcommand{\aaa}{\BoldVec{a} {}}

\newcommand{\jj}{\BoldVec{j} {}}

\newcommand{\KK}{\BoldVec{K} {}}

\newcommand{\GG}{\BoldVec{G} {}}
\newcommand{\kk}{\BoldVec{k} {}}

\newcommand{\nab}{\BoldVec{\nabla} {}}

\newcommand{\oo}{\BoldVec{\omega} {}}

\newcommand{\ii}{{\rm i}}

\newcommand{\dd}{{\rm d} {}}

\def\half{{\textstyle{1\over2}}}

\def\onethird{{\textstyle{1\over3}}}
\def\twothird{{\textstyle{2\over3}}}
\def\fourthird{{\textstyle{4\over3}}}

\newcommand{\yan}[3]{, Astron. Nachr. {\bf #2}, #3 (#1).}

\newcommand{\yana}[3]{, Astron. Astrophys. {\bf #2}, #3 (#1).}

\newcommand{\yjfm}[3]{, J. Fluid Mech. {\bf #2}, #3 (#1).}

\newcommand{\ypre}[3]{, Phys.\ Rev.\ E \ {\bf #2}, #3 (#1).}
\newcommand{\yprl}[3]{, Phys.\ Rev.\ Lett.\ {\bf #2}, #3 (#1).}

\newcommand{\yapj}[3]{, Astrophys. J. {\bf #2}, #3 (#1).}

\newcommand{\ygafd}[3]{, Geophys. Astrophys. Fluid Dynam. {\bf #2}, #3 (#1).}

\newcommand{\yjour}[4]{, #2 {\bf #3}, #4 (#1).}

\newcommand{\ybook}[3]{, {\em #2}. #3 (#1).}

\begin{document}

\preprint{NORDITA 2004-64}

\title{Nonlinear current helicity fluxes in turbulent dynamos
and alpha quenching}
\author{Kandaswamy Subramanian}
\affiliation{IUCAA, Post bag 4, Ganeshkhind, Pune 411 007, India}
\email{kandu@iucaa.ernet.in}
\author{Axel Brandenburg}
\affiliation{NORDITA, Blegdamsvej 17, DK-2100 Copenhagen \O, Denmark}
\email{brandenb@nordita.dk}

\date{\today}
\begin{abstract}

Large scale dynamos produce small 
scale current helicity as a waste product that quenches the large 
scale dynamo process (alpha effect). This quenching can be
catastrophic (i.e.\ intensify with magnetic Reynolds number) 
unless one has fluxes of small scale
magnetic (or current) helicity out of the system.
We derive the form of helicity fluxes in 
turbulent dynamos, taking also into account the nonlinear effects of
Lorentz forces due to fluctuating fields. We confirm the form
of an earlier derived magnetic helicity flux term,
and also show that
it is not renormalized by the small scale magnetic field,
just like turbulent diffusion. Additional
nonlinear fluxes are identified, which are driven by
the anisotropic and antisymmetric parts of the magnetic correlations.
These could provide further ways for turbulent dynamos
to transport out small scale magnetic helicity, so as
to avoid catastrophic quenching.

\end{abstract}

\maketitle

Large scale magnetic fields in astrophysical 
bodies are thought to be generated by dynamo action involving 
helical turbulence and rotational shear \cite{dynam,Mof78}. 
A particularly important driver of the mean field dynamo (MFD)
is the $\alpha$ effect, which in the kinematic regime
is proportional to the kinetic helicity of the turbulence.
A question of considerable debate is how the $\alpha$ effect gets 
modified due to the backreaction of
the generated mean and fluctuating fields? It is especially
important to understand whether $\alpha$
suffers catastrophic (i.e.\ $R_{\rm m}$-dependent) quenching,
since $R_{\rm m}$, the magnetic Reynolds number, is expected to be
typically very large in astrophysical systems.
Recent progress has come from realizing the importance of
magnetic helicity conservation in constraining this nonlinear
saturation \cite{GD,axks04}.

Recall that in the MFD theory,
one splits the magnetic field $\BB$ into a mean
magnetic field $\meanBB$ and a small scale
field  $\bb =\BB - \meanBB$, and
derives the mean-field dynamo equation \cite{dynam}
\EQ
\frac{\partial \meanBB}{\partial t} =
\nab \times \left( \meanUU \times \meanBB + \meanemf -
\eta\meanJJ \right).
\label{basicm}
\EN
Here $\meanemf\equiv\overline{\uu\times\bb}$ is the turbulent
electromotive force (emf), $\meanJJ=\nab\times\meanBB/\mu_0$
the mean current density, $\mu_0$ the vacuum permeability
(assumed unity throughout the rest of the paper),
$\eta$ the microscopic resistivity and the velocity 
$\UU = \meanUU + \uu$ has also
been split into mean $\meanUU$ and small scale
turbulent $\uu = \UU - \meanUU$ velocities.
Finding an expression for the correlator $\meanemf$ in terms of the
mean fields is a standard closure problem which is at the heart of
mean field theory. In the two-scale approach \cite{Mof78} one assumes that
$\meanemf$ can be expanded in powers of the gradients of the mean
magnetic field. For isotropic helical turbulence, this gives
$\meanemf \approx \alpha \meanBB - \eta_{\rm t}\meanJJ$,
where in the kinematic limit, 
$\alpha = \alpha_{\rm K} = -\onethird\tau\overline{\oo\cdot\uu}$, 
proportional to the kinetic helicity ($\oo=\nab\times\uu$) and 
$\eta_{\rm t} = \onethird\tau\overline{\uu^2}$
is the turbulent magnetic diffusivity proportional to the
specific kinetic energy of the turbulence, with
$\tau$ being the velocity correlation time.

The kinematic theory has to be modified to take account
of the backreaction due to the Lorentz forces associated
with the generated large and small scale fields.
Treatments of the back-reaction  
have typically used the quasi-linear approximation 
or closure schemes to derive corrections to the mean-field dynamo 
coefficients (cf.\ \cite{pouq,GD} and \cite{axks04} for a review). 
It is then found that the $\alpha$ effect gets 
``renormalized'' by the addition of a term proportional to the 
current helicity, $\overline{\jj\cdot\bb}$, of the small scale fields;
that is $\alpha=\alpha_{\rm K} + \alpha_{\rm M}$, where 
$\alpha_{\rm M} = (\tau/3\rho_0)\overline{\jj\cdot\bb}$ \cite{pouq}.
(Here $\rho_0$ is the density of the fluid and $\jj$ the small
scale current density.)
At the same time there is no modification to $\eta_{\rm t}$ at lowest order \cite{GD}.

In order to constrain $\alpha_{\rm M}$, in simulations in a periodic box
or in systems that involve no boundaries, it has proved useful
to take recourse to the
evolution equation for the small scale magnetic helicity
$h=\bra{\overline{\aaa\cdot\bb}}$, where $\aaa$ is the vector potential 
and $\bra{\,}$ denotes volume averaging.  We have
in such situations,
$\dd h/\dd t= -2\bra{\meanemf\cdot\meanBB}-2\eta\bra{\overline{\jj\cdot\bb}}$.
In the stationary limit, this predicts $\bra{\meanemf\cdot\meanBB}
= -\eta\bra{\jj\cdot\bb}$, which tends to zero as $\eta \to 0$,
for any reasonable spectrum of current helicity. This leads to a catastrophic
quenching of the turbulent emf parallel to $\meanBB$. 
Of course while evolving to this stationary state, some $\meanBB$ will
be generated. But its value, at the end of the kinematic regime, 
still turns out to be small, if 
the scale separation is large \cite{dynquench}.
It has been suggested that such quenching can be avoided if 
the system has open boundaries and is 
inhomogeneous, since one could then have a flux of small scale
helicity out of the system which helps maintain a non-zero 
$\bra{\meanemf\cdot\meanBB}$ \cite{bf01,klee,VC01}. 
It is therefore important to
calculate such fluxes in a general manner, taking into account
also the effect of Lorentz forces. This is the aim of the present work.

Indeed, Vishniac and Cho \cite{VC01} derived an 
interesting flux of helicity,
which arises even for nonhelical but anisotropic turbulence.
We derive a generalized form of the Vishniac-Cho flux (henceforth VC flux)
in the evolution equation for $\overline{\jj\cdot\bb}$
to include also nonlinear effects of the Lorentz force
and helicity in the fluid turbulence.
As we see, the VC flux can also be thought of as
a generalized anisotropic turbulent diffusion.
Further, due to nonlinear effects, other helicity flux contributions arise 
generated by the anisotropic and antisymmetric part of the 
magnetic correlations. 

One immediate problem with previous approaches was that 
in open systems with boundaries,
$h$ is not gauge invariant and one has to consider instead
the gauge-invariant relative magnetic helicity, say $h_R$,
defined by subtracting the helicity of a reference vacuum field \cite{relhel}.
The flux of relative helicity is cumbersome
to work with for arbitrarily shaped boundaries. 
Also the concept of a density of relative helicity is
not meaningful, since $h_R$ is defined only
as a volume integral. In order to avoid these problems
it is advantageous to consider instead the current helicity
$\overline{\jj\cdot\bb}$ and its flux. The current helicity
density and its flux are directly gauge invariant, locally
well defined, and are in fact the observationally measured 
quantities in say the solar context. Furthermore it is $\overline{\jj\cdot\bb}$
which directly enters the expression for the nonlinear 
$\alpha$ effect \cite{pouq,GD}. Also for isotropic small scale
fields the spectra of small scale current ($C_k$) 
and magnetic helicities ($H_k$) are related by $H_k = C_k/k^2$.
For these reasons we consider here directly the current helicity
evolution and use the current helicity flux as a `proxy' for
the magnetic helicity flux.

\noindent{\it Current helicity evolution}.
Consider the evolution of the small scale current helicity,
$\overline{\jj\cdot\bb} = \epsilon_{ijk}\overline{b_i\partial_jb_k}$,
which is explicitly gauge invariant.
We assume that the correlation tensor of 
fluctuating quantities ($\uu$ and $\bb$) vary slowly on
the system scale, say $\RR$. Consider the equal time,
ensemble average of the product $\overline{f(\xx_1)g(\xx_2)}$.
The common dependence of $f$ and $g$ on $t$ is assumed
and will not explicitly be stated.
Let  $\hat{f}(\kk_1)$ and $\hat{g}(\kk_2)$ be the Fourier transforms
of $f$ and $g$, respectively.
We can express this correlation as
$\overline{f(\xx_1)g(\xx_2)}
= \int \Phi(\hat{f},\hat{g},\kk,\RR) \ {\rm e}^{i\kk\cdot\rrr} \
\dd^3k$, with 
\EQ
\Phi(\hat{f},\hat{g},\kk,\RR)
= \int \overline{\hat{f}(\kk + \half\KK) \hat{g}(-\kk +\half\KK)}
\ {\rm e}^{i\KK\cdot\RR} \,\dd^3K .
\label{phi_def}
\EN
Here we have defined the difference $\rrr = \xx_1 - \xx_2$ and the mean
$\RR = \half(\xx_1 + \xx_2)$, keeping in mind that all
two-point correlations will vary rapidly with $\rrr$ but slowly
with $\RR$ \cite{RS75}. Also
$\kk = \half(\kk_1 - \kk_2)$ and $\KK = \kk_1 + \kk_2$.
In what follows we require the correlation
tensors, $v_{ij}(\kk,\RR) = \Phi(\hat{u}_i,\hat{u}_j,\kk,\RR)$, 
of the $\uu$  field, $m_{ij}(\kk,\RR) = \Phi(\hat{b}_i,\hat{b}_j,\kk,\RR)$
of the $\bb$ field and the cross correlation
$\chi_{jk}(\kk,\RR) = \Phi(\hat{u}_j,\hat{b}_k,\kk,\RR)$
between these two fields, in Fourier space. The turbulent emf
is given by $\meanemfs_i(\RR) = \epsilon_{ijk} \int \chi_{jk}(\kk,\RR) 
\,\dd^3k$.

In order to compute the current helicity 
evolution of $\partial\overline{\jj(\xx)\cdot\bb(\xx)}/\partial t$, 
we use the induction equation for $\bb$ in Fourier space,
\EQA
\frac{\partial{\hat{b}}_k(\kk)}{\partial t}
&=&\epsilon_{kpq}\epsilon_{qlm} \ii k_p \int \hat{u}_l(\kk - \kk')
\hat{\meanB}_m(\kk') \,\dd^3k'
\nonumber \\
&& +\hat{G}_k(\kk) -\eta k^2 \hat{b}_k(\kk).
\label{bhat}
\ENA
Here, $\GG = \nab\times(\uu\times\bb - \overline{\uu\times\bb})$
is the nonlinear term. (We also neglect the 
velocity shear due to $\UU$ compared to that due to $\uu$.)
A tedious but straightforward calculation gives \cite{axks04}
\EQA
{\partial \over \partial t}\,\overline{\jj\cdot\bb}
&&= \epsilon_{ljk}\int \Big[ 2\chi_{lk} \  k_j (\kk\cdot\meanBB)
-\chi_{lk} \nabla_j(\ii \kk\cdot\meanBB)
\nonumber \\
&&
-\ii k_j \ \meanBB\cdot\nab\chi_{lk}
+ 2 \ii k_j \chi_{pk} \nabla_p\meanB_l \Big]\,\dd^3k
+T_C.
\label{curhelfin}
\ENA
We have written out explicitly only that part
of the helicity evolution driven by the coupling of the 
turbulent emf to the mean magnetic field.
This is because we are particularly interested in
turbulent helicity fluxes driven by an inhomogeneous 
$\meanemf$ and $\meanBB$.
The $T_C$ term represents the triple correlations of the small scale
$\uu$ and $\bb$ fields and the microscopic diffusion terms that one gets 
on using \Eq{bhat}. The handling of the triple correlations
needs a closure approximation.  But we will not
need to explicitly evaluate these terms to
identify the helicity fluxes we are interested in; i.e.\
those which couple $\meanemf$ and $\meanBB$,
and so continue to write this term as $T_C$.

In order to calculate the current helicity evolution, using
\Eq{curhelfin}, we have to calculate also $\chi_{lk}$. 
This has been done in detail by \cite{RKR03,axks04}. 
(One adopts a closure approximation whereby
triple correlations, $T_{lk}$, which
arise now in the evolution equation for $\partial\chi_{lk}/\partial t$,
are assumed to provide relaxation of the turbulent emf 
or $\chi_{lk}$ and one takes $T_{lk} = -\chi_{lk}/\tau$, where
$\tau$ is the relaxation time (cf. also \cite{BF02})).
We concentrate
below on nonrotating but helical turbulence.
For such turbulence we have from \cite{RKR03,axks04},
$\chi_{lk} = \tau {\sf I}_{lk}$, where ${\sf I}_{lk}$
is given by; see Eq.~(10.30) in \cite{axks04}. 
\EQA
{\sf I}_{lk} &=& 
-\ii\kk\cdot\meanBB(v_{lk}- m_{lk})
+ \half\meanBB\cdot\nab(v_{lk} + m_{lk})
\nonumber \\
&&+ \meanB_{l,s} m_{sk}
- \meanB_{k,s} v_{ls}
- \half k_m \ \meanB_{m,s}
\left({\partial v_{lk} \over \partial k_s}
+ {\partial m_{lk} \over \partial k_s}\right)
\nonumber \\
&&- 2{k_lk_s \over k^2} \meanB_{s,p}m_{pk}.
\label{Ilk}
\ENA
We use this in what follows.

Let us denote the four terms under the integral
in \Eq{curhelfin} by $A_1$, $A_2$, $A_3$ and $A_4$, respectively.
In $A_1$, due to the presence of $\epsilon_{ljk}$, only
the antisymmetric parts of the tensors $v_{lk}$ and $m_{lk}$
survive, and these are denoted by $v_{lk}^{\rm A}$ and $m_{lk}^{\rm A}$,
respectively. Also note that the last term above
vanishes because it involves the product $\epsilon_{ljk}k_lk_j =0$.
All the other terms of \Eq{curhelfin} already have one $R$ derivative,
and so one only needs to retain the term in $\chi_{lk} = \tau 
{\sf I}_{lk}$ which does not contain $R$ derivatives. 
In $A_3$ one can then use the fact that
$\nab\cdot\meanBB =0$ to write it as a total divergence.
We now turn to specific cases.

\noindent{ \it Isotropic, helical, nonrotating turbulence}.
Let us first reconsider the simple case of 
isotropic, helical, nonrotating, and weakly inhomogeneous turbulence. 
For such turbulence, the form of the velocity and magnetic correlation
tensors is given by \cite{RKR03,RS75}.
In evaluating the $k$-integrals, only terms which involve integration
over an even number of $k_i$ survive. Also, in terms which
already involve one $R_i$ derivative, one needs to keep only
the homogeneous terms in $v_{lk}$ and $m_{lk}$. Further in the presence of
$\epsilon_{ljk}$ all terms symmetric in any pair of the indices vanish.
Taking account of these considerations, it turns out that only the
homogeneous part of the velocity and magnetic correlation tensors
given in \cite{RKR03,RS75} survive. The homogeneous part
of these correlations is
\EQ
v_{ij} = \left[\delta_{ij} - {k_ik_j\over k^2}\right] E(k,\RR) 
- {\epsilon_{ijk} \ii k_k \over k^2} F(k,\RR),
\label{vel_cor}
\EN
and a similar expression for $m_{ij}$ with functions say $M(k,\RR)$
and $N(k,\RR)$ replacing $E$ and $F$, respectively.
Here $4\pi k^2E$ and $4\pi k^2M$ are
the kinetic and magnetic energy spectra, respectively,
and $4\pi k^2F$ and $4\pi k^2N$ are the corresponding helicity spectra.
They obey the relations
$\overline{\uu^2} = 2 \int E\dd^3k$,
$\overline{\uu\cdot\nab\times\uu} = 2 \int F\dd^3k$,
$\overline{\bb^2} = 2 \int M\dd^3k$, and 
$\overline{\jj\cdot\bb} = 2 \int N\dd^3k$.
With these simplifications we have, after carrying out the angular 
integrals over the unit vectors $\kunit_i =k_i/k$,
\EQ
A_1 = \fourthird\meanBB^2 \int k^2(F-N)\,\dd^3k 
+\twothird\meanBB\cdot\meanJJ \int k^2 (M+E)\,\dd^3k.
\EN

In the case of isotropic turbulence, the second and third terms,
$A_2$ and $A_3$, are zero
because, to leading order in $R$ derivatives,
the integrands determining $A_2$ and $A_3$ have an odd number 
(3) of $\kunit_i$'s. The fourth term is given by 
\EQ
A_4 = \twothird\meanJJ\cdot\meanBB \tau\int k^2 [E-M]\,\dd^3k.
\EN
Adding all the contributions, $A_1+A_2+A_3+A_4$, we get
for the isotropic, helical, weakly inhomogeneous turbulence,
\EQ
{\partial \over \partial t}\,\overline{\jj\cdot\bb} =
\fourthird\meanBB^2 \tau\int k^2(F-N)\,\dd^3k 
+\fourthird\meanJJ\cdot\meanBB \tau \int k^2 E\,\dd^3k
+T_C.
\label{curisofin}
\EN
We see that there is a nonlinear correction due to the
small scale helical part of the magnetic correlation to
the term $\propto \meanBB^2$. But the nonlinear
correction to the term $\propto \meanJJ\cdot\meanBB$
has canceled out, just as there is no such
correction to turbulent diffusion \cite{GD}. 

As pointed out above, in the isotropic case
the magnetic helicity spectrum is $H_k = C_k/k^2$.
So the first two terms of the current
helicity evolution equation \Eq{curisofin} can be interpreted
as representing the effects
of exactly the source term $-2\meanemf\cdot\meanBB$ which is obtained
for the magnetic helicity evolution. 
Also for this isotropic, but weakly
inhomogeneous case, one sees that
there is no flux which explicitly depends on the mean magnetic field.

\noindent{\it Anisotropic turbulence}.
Let us now consider anisotropic turbulence.
In the first term $A_1$ in \Eq{curhelfin},
one cannot now assume the isotropic form for the
velocity and magnetic correlations.
But again, due to the presence of $\epsilon_{ljk}$, only
the antisymmetric parts of the tensors $v_{lk}$ and $m_{lk}$
survive. Also the last term in \Eq{Ilk} does not contribute to $A_1$ 
because it involves the product $\epsilon_{ljk}k_lk_j =0$.
One can further simplify the term involving $k$ derivatives
by integrating it by parts. Straightforward algebra, and a 
judicious combination of the terms then gives
\EQA
A_1&=&
\tau\epsilon_{ljk} \Big\{
 -2\ii \meanB_p\meanB_s \int k_j k_p k_s 
(v_{lk}^{\rm A}- m_{lk}^{\rm A})\,\dd^3k
\nonumber \\
&& +2 \meanB_p \int k_j k_p (  
\meanB_{l,s} m_{sk} - \meanB_{k,s} v_{ls})\,\dd^3k
\nonumber \\
&&+ \meanB_p \meanB_{m,j} \int k_m k_p (
v_{lk}^{\rm A} +  m_{lk}^{\rm A})\,\dd^3k
\nonumber \\
&& + \nabla_s \left[\meanB_p\meanB_s \int k_j k_p
(v_{lk}^{\rm A} +  m_{lk}^{\rm A}) \right] \Big\}\,\dd^3k.
\label{A1aniso}
\ENA
All the other terms, $A_2$, $A_3$ and $A_4$, cannot be further simplified.
They are explicitly given by
\EQ
A_2= -\tau\epsilon_{ljk}\meanB_p \meanB_{s,j} \int k_s k_p
(v_{lk}^{\rm A}- m_{lk}^{\rm A})\,\dd^3k,
\label{A2aniso}
\EN
\EQ
A_3 = -\nabla_s\left[\tau\epsilon_{ljk} \meanB_p \meanB_s 
\int k_jk_p (v_{lk}^{\rm A}- m_{lk}^{\rm A}) \right]\,\dd^3k,
\label{A3aniso}
\EN
\EQ
A_4 =
2\tau\epsilon_{ljk} \meanB_p \meanB_{l,s} \int k_j k_p 
(v_{sk}- m_{sk})\,\dd^3k.
\label{A4aniso}
\EN
Adding all the contributions, $A_1+A_2+A_3+A_4$, we get
\EQA
&&{\partial \over \partial t}\,\overline{\jj\cdot\bb} =
2\epsilon_{jlk}\tau \Big[
\meanB_p\meanB_s \int \ii k_j k_p k_s
(v_{lk}^{\rm A}- m_{lk}^{\rm A})\,\dd^3k
\nonumber \\
&& 
+ 2 \meanB_p\meanB_{k,s} \int k_j k_p v_{ls}^{\rm S}\,\dd^3k
-\meanB_p \meanB_{s,j} \int k_s k_p m_{lk}^{\rm A}\,\dd^3k
\nonumber \\
&& 
-\nabla_s \left(\meanB_p \meanB_s
\int k_jk_p  m_{lk}^{\rm A} \right) \,\dd^3k \Big] + T_C.
\label{anisochel}
\ENA
Here $v_{ls}^{\rm S} = \half(v_{ls} + v_{sl})$ is the symmetric part of
the velocity correlation function.

Let us discuss the various effects contained in 
\Eq{anisochel} for current helicity evolution.
The first term in \Eq{anisochel} represents the anisotropic
version of helicity generation due to the full nonlinear $\alpha$ effect.
In fact, for isotropic turbulence it exactly will match the
first term in \Eq{curisofin}. The second term in
\Eq{anisochel} gives the effects on helicity evolution
due to a generalized anisotropic turbulent diffusion.
This is the term which contains the VC flux.
To see this, rewrite this term as
\EQA
&&\left.{\partial\overline{\jj\cdot\bb} \over \partial t}\right\vert_V =
4\tau \epsilon_{jlk} \meanB_p\meanB_{k,s} \int k_j k_p v_{ls}^{\rm S}\,\dd^3k
\nonumber \\
&=& -\nab\cdot\meanFF^V
+4\tau\meanB_k \epsilon_{klj} \meanB_{p,s} \int k_j k_p v_{ls}^{\rm S}\,\dd^3k.
\label{vischoa}
\ENA
Here the first term is the VC flux,
$\meanF^V_s=\phi_{spk}\meanB_p\meanB_k$, where
$\phi_{spk}$ is a new turbulent transport tensor with
\EQ
\phi_{spk}
=-4\tau\epsilon_{jlk} \int k_j k_p v_{ls}^{\rm S}\,\dd^3k
=-4\tau\overline{\omega_k \nabla_p u_s}.
\label{vishflux}
\EN
Obviously, only components of $\phi_{spk}$ symmetric in
$p$ and $k$ enters in the flux $\meanF^V_s$.
The second term in \Eq{vischoa} is the effect on helicity due to `anisotropic
turbulent diffusion'. (We have not included the large scale
derivative of $v_{ls}$ to the leading order.)
Strictly speaking, $\calFF^V$ is a current helicity flux,
but if we define the spectrum of the
magnetic helicity flux by dividing 
the spectrum of the current helicity flux by a $k^2$ factor,
\Eq{vishflux} for $\calFF^V$ leads
exactly to the magnetic helicity flux given 
in Eqs.~(18) and (20) of Vishniac and Cho \cite{VC01}.

This split into helicity flux and anisotropic diffusion 
may seem arbitrary; some support for its usefulness comes
from the fact that, for isotropic turbulence, 
$\calFF^V$ vanishes, while the second
term exactly matches with the corresponding
helicity generation due to turbulent diffusion, i.e.\
the $\meanJJ\cdot\meanBB$ term in \Eq{curisofin}.
Of course, we could have just retained the non-split expression in 
\Eq{vischoa}, which can then be looked at
as an effect of anisotropic turbulent
diffusion on helicity evolution. 
Also, interestingly, there is no nonlinear
correction to the VC flux
from the small scale magnetic field in the form of,
say, a term proportional to $m_{ls}^{\rm S}$;
just as previously, there was no
nonlinear correction to turbulent diffusion in lowest order!

Finally, \Eq{anisochel} also contains terms (the last two)
involving only the antisymmetric parts of
the magnetic correlations. These terms
vanish for isotropic turbulence, but contribute
to helicity evolution for nonisotropic turbulence.
The last term gives a purely magnetic contribution
to the helicity flux, but one that depends only
on the antisymmetric part of $m_{lk}$. 
Note that such magnetic correlations, even
if initially small, may spontaneously develop
due to the kinematic $\alpha$ effect or anisotropic turbulent diffusion
and may again provide a helicity flux. More work
is needed to understand this last flux term
better \cite{KR99,KleeFlux}.
Preliminary simulations of helical turbulence with shear and open
boundaries suggest that the sign of the VC flux agrees
with that of the small scale current helicity flux, but that its magnitude
may only account for about 25\% of the actual flux \cite{BranSan04}.
The existence of the VC flux has also been verified in simulations of
nonhelically driven shear flow turbulence \cite{AB01}, but its
magnitude was too small to produce dynamo action.

In conclusion, we have derived helicity fluxes in 
turbulent dynamos, taking also into account the nonlinear effects of 
Lorentz forces due to the fluctuating field. To avoid gauge ambiguities,
we have followed the current helicity evolution.
We confirm the form
of the helicity flux found by Vishniac and Cho, who used
the first order smoothing approximation. 
We note however that
it is more correctly interpreted as a current helicity flux
and not as a flux of relative magnetic helicity.
In addition we have 
found that the corresponding turbulent coefficient
does not get renormalized due to nonlinear effects,
just as is the case of turbulent diffusion. Additional
nonlinear fluxes have been identified as being driven by
the anisotropic and antisymmetric parts of the magnetic correlations.
These could provide further ways for turbulent dynamos
to transport out small scale magnetic helicity so as
to avoid catastrophic $R_{\rm m}$-dependent quenching.
It remains to calculate these fluxes in specific circumstances
and also verify their presence in direct numerical
simulations of turbulent dynamos.
\begin{acknowledgments}
KS thanks NORDITA for hospitality during the course of this work.
\end{acknowledgments}

\vfill\bigskip\noindent{\it
$ $Id: paper.tex,v 1.29 2004/09/29 14:12:15 brandenb Exp $ $}


\begin{thebibliography}{99}
\bibitem{dynam} Ya. B. Zeldovich, A. A. Ruzmaikin, and D. D. Sokoloff, 
{\it Magnetic fields in Astrophysics}, Gordon and Breach, New York (1983);
A. A. Ruzmaikin, A. M. Shukurov, and D. D. Sokoloff, {\it Magnetic fields
of galaxies}, Kluwer, Dordrecht (1983);
E. Parker, {\it Cosmic magnetic fields}, Clarendon
, Oxford (1979), R. Beck, A. Brandenburg, D. Moss, A. Shukurov,
and D. Sokoloff, Ann. Rev. Astron. Astrophys., {\bf 34}, 155 (1996).
\bibitem{Mof78}
H. K. Moffatt\ybook{1978}
{Magnetic Field Generation in Electrically Conducting Fluids}
{Cambridge University Press, Cambridge}
\bibitem{axks04} A. Brandenburg and K. Subramanian, Phys. Rept.
(submitted), e-print astro-ph/0405052.
\bibitem{GD} A. V. Gruzinov and P. H. Diamond, Phys. Rev. Lett.,
{\bf 72}, 1651 (1994); K. Avinash, Phys. Fluids B, {\bf 3}, 2150 (1991);
K. Subramanian\yprl{2003}{90}{245003} 
\bibitem{pouq} A. Pouquet, U. Frisch, and J. L\'eorat, 
J. Fluid Mech., {\bf 77}, 321 (1976).
\bibitem{dynquench} E. Blackman and A. Brandenburg, 
Astrophys. J., {\bf 579}, 359 (2002);
K. Subramanian, Bull. Astr. Soc. India, {\bf 30}, 715 (2002).
\bibitem{VC01} E. T. Vishniac and J. Cho\yapj{2001}{550}{752}
\bibitem{bf01} E. G. Blackman and G. F. Field, Phys. Plasmas,
{\bf 8}, 2407 (2001).
\bibitem{klee}
N. I. Kleeorin, D. Moss, I. Rogachevskii, D. Sokoloff\yana{2000}{361}{L5};
{\bf 387}, 453 (2002); {\bf 400}, 9 (2003).
\bibitem{relhel}
M. Berger and G. B. Field\yjfm{1984}{147}{133}; 
J. M. Finn and T. M. Antonsen\yjour{1985}{Comments Plasma Phys.\ Controlled
Fusion}{9}{111}
\bibitem{RS75}
P. H. Roberts and A. M. Soward\yan{1975}{296}{49}
\bibitem{RKR03}
K.-H. R\"adler, N. Kleeorin, and I. Rogachevskii\ygafd{2003}{97}{249}
\bibitem{BF02}
E. G. Blackman, G. B. Field\yprl{2002}{89}{265007}
\bibitem{KR99} N. I. Kleeorin and I. Rogachevskii\ypre{1999}{59}{6724}
\bibitem{KleeFlux}Note that our
helicity flux is distinct from that discussed
by Kleeorin and coworkers \cite{klee}.
In the original treatment involving path integral methods,
a flux proportional to the 
anisotropic part of the $\alpha$ tensor and the anisotropic
part of the magnetic helicity tensor were derived \cite{KR99}.
However, the subsequent use of this flux in \cite{klee}
involved adopting a phenomenological term proportional to
the mean field; such a term should have arisen automatically
in our treatment.
\bibitem{BranSan04}
A. Brandenburg and C. Sandin, Astron. Astrophys. (in press),
e-print astro-ph/0401267.
\bibitem{AB01}
R. Arlt and B. Brandenburg\yana{2001}{380}{359}

\end{thebibliography}
\end{document}